\documentclass[amsmath,amssymb,aps,prl,reprint,superscriptaddress]{revtex4-2}

\usepackage{graphicx}% Include figure files
\usepackage{dcolumn}% Align table columns on decimal point
\usepackage{bm}% bold math
\usepackage{tikz}
\usepackage{hyperref}
\usetikzlibrary{calc}
\usepackage{braket}
\usetikzlibrary{fit,backgrounds}

\begin{document}

\preprint{APS/123-QED}

\title{Combining matrix product states and mean-field theory to capture magnetic order in quasi-1D cuprates\\}% Force line breaks with \\

\author{Q. Staelens}
\author{D. Verraes}
\author{D. Vrancken}
\affiliation{Center for Molecular Modeling, Ghent University, 9052 Zwijnaarde, Belgium}
\affiliation{Department of Physics and Astronomy, Ghent University, 9000 Ghent, Belgium}
\author{T. Braeckevelt}
\affiliation{Center for Molecular Modeling, Ghent University, 9052 Zwijnaarde, Belgium}

\author{J. Haegeman}
\affiliation{Department of Physics and Astronomy, Ghent University, 9000 Ghent, Belgium}

\author{V. Van Speybroeck}
\affiliation{Center for Molecular Modeling, Ghent University, 9052 Zwijnaarde, Belgium}

\date{\today}% It is always \today, today,
             %  but any date may be explicitly specified

\begin{abstract}
We study quasi-one-dimensional strongly correlated materials using a multi-step approach based on density functional theory, downfolding techniques, and tensor-network simulations.
The downfolding procedure yields effective multiband Hubbard models that capture the competition between electron hopping and local Coulomb interactions relevant to the system’s low-energy properties.
The resulting multiband Hubbard models are solved using matrix product states.
Applied to \(\text{Sr}_2\text{CuO}_3\), \(\text{SrBaCuO}_3\), and \(\text{Ba}_2\text{CuO}_3\), this purely one-dimensional treatment yields no long-range magnetic order, in contrast to the magnetic ordering observed experimentally.
To account for this behavior, we extend the multi-step approach by incorporating interchain couplings through a self-consistent mean-field scheme.
This combined approach stabilizes finite staggered magnetizations, providing a consistent description of magnetic order in agreement with experiment.
For \(\text{Sr}_2\text{CuO}_{3.5}\) and \(\text{SrCuO}_2\), we also tested an approach proposed for ladder materials, however, we find that these materials are not well suited for this approach due to the small magnitude of the intraladder hopping parameters.
\end{abstract}

%\keywords{Suggested keywords}%Use showkeys class option if keyword
                              %display desired
\maketitle

%\tableofcontents

\section{\label{sec:level1}I. Introduction}

In many-body quantum systems, directly solving the full Hamiltonian is often computationally intractable due to the exponential growth of the Hilbert space with system size. While density functional theory (DFT) \cite{Hohenberg1964, Kohn1965} has become a standard and highly successful tool for describing the ground-state properties of weakly correlated materials, it encounters fundamental limitations in the presence of strong electronic correlations \cite{Cohen2008, Morosan2012}. In particular, commonly used local and semilocal exchange–correlation functionals struggle to accurately describe systems where electron-electron interactions dominate over kinetic energy, leading to qualitative failures such as the incorrect prediction of metallic behavior in Mott insulators \cite{Imada1998}, the underestimation of band gaps, and the inability to capture local magnetic moments. These shortcomings indicate that an accurate description of strongly correlated materials requires an explicit treatment of low-energy electronic interactions beyond effective single-particle theories \cite{Jiang2015}.

To address this challenge, one powerful strategy is downfolding the system to an effective Hamiltonian. In this approach, a complex, high-dimensional Hamiltonian is systematically reduced to a simpler, low-energy effective Hamiltonian that captures the essential physics of the strongly correlated electrons \cite{Imada2010}.
The screening of the bare Coulomb interaction by electrons outside the low-energy subspace should be incorporated into the effective two-body interaction parameters, whereas screening processes involving low-energy electrons must be excluded, as they are already treated explicitly within the effective Hamiltonian. 
Several constrained approaches have been proposed to tackle this issue \cite{Mahan1988, Hirayama2013}. Among them, the constrained random phase approximation (cRPA) \cite{Aryasetiawan2004, Aryasetiawan2006} has become one of the most widely used methods for evaluating screened interaction parameters in solids, largely because it can naturally capture the frequency dependence of the interactions \cite{Biermann2014}.
This approach yields a screened Coulomb interaction which, combined with the kinetic term describing electron hopping between orbitals, defines an effective Hamiltonian taking the form of an extended multiband Hubbard model \cite{Hubbard1963}. After constructing the Hubbard model, the next challenge is to solve it. 

Dynamical mean-field theory (DMFT) \cite{Georges1996, Kotliar2006} maps the Hubbard model to an Anderson impurity model \cite{Anderson1961}, which can be solved with high precision using methods such as quantum Monte Carlo (QMC) \cite{Zhang2003} or coupled-cluster (CC) techniques \cite{Bartlett2007}. Unfortunately, QMC methods are limited by the sign problem \cite{Loh1990}, while coupled-cluster approaches become impractical for realistic multiband systems. 
In contrast, tensor-network (TN) \cite{Cirac2021} techniques have been developed as efficient tools for interacting quantum systems \cite{Chan2011, Fertitta2014, Baiardi2020}. The density matrix renormalization group (DMRG) \cite{White1992, White1993, Schollwock2011} on matrix product states (MPS) remains highly effective in one dimension (1D), while projected entangled-pair states (PEPS) \cite{Verstraete2004, Cirac2021} are advancing for higher-dimensional and fermionic systems, though research is still ongoing to enhance their efficiency and scope. Recently, our group employed TN methods in combination with downfolding techniques to investigate strongly correlated materials \cite{Vrancken2025}.

Motivated by the performance of TNs in 1D, we therefore focus on quasi-1D strongly correlated materials. In 1D at $T=0$ K, it is impossible to spontaneously break the continuous \( SU(2) \) spin rotation symmetry. This is better known as the Mermin-Wagner-Hohenberg theorem \cite{Mermin1966, Pitaevskii1991, Tanaka2004}. It is nevertheless clear that in any real material, such as Sr$_2$CuO$_3$ \cite{Ami1995}, purely one-dimensional behavior is only approximate, as some degree of interchain coupling is present. As a result, three-dimensional (3D) magnetic long-range order can emerge below a finite N\'eel temperature $T_N$. In this work, we show that a conceptually simple treatment of the interchain coupling within a mean-field (MF) approximation provides a coherent description of the ordered phase and yields nontrivial quantitative predictions for static properties.

\begin{figure*}
\centering
\begin{tikzpicture}
\node[anchor=south west] at (-2,0) {
  \includegraphics[width=0.1\textwidth]{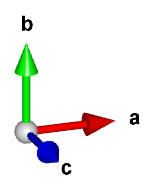}
};

\node[anchor=south west] at (0,0) {
  \includegraphics[width=0.75\textwidth]{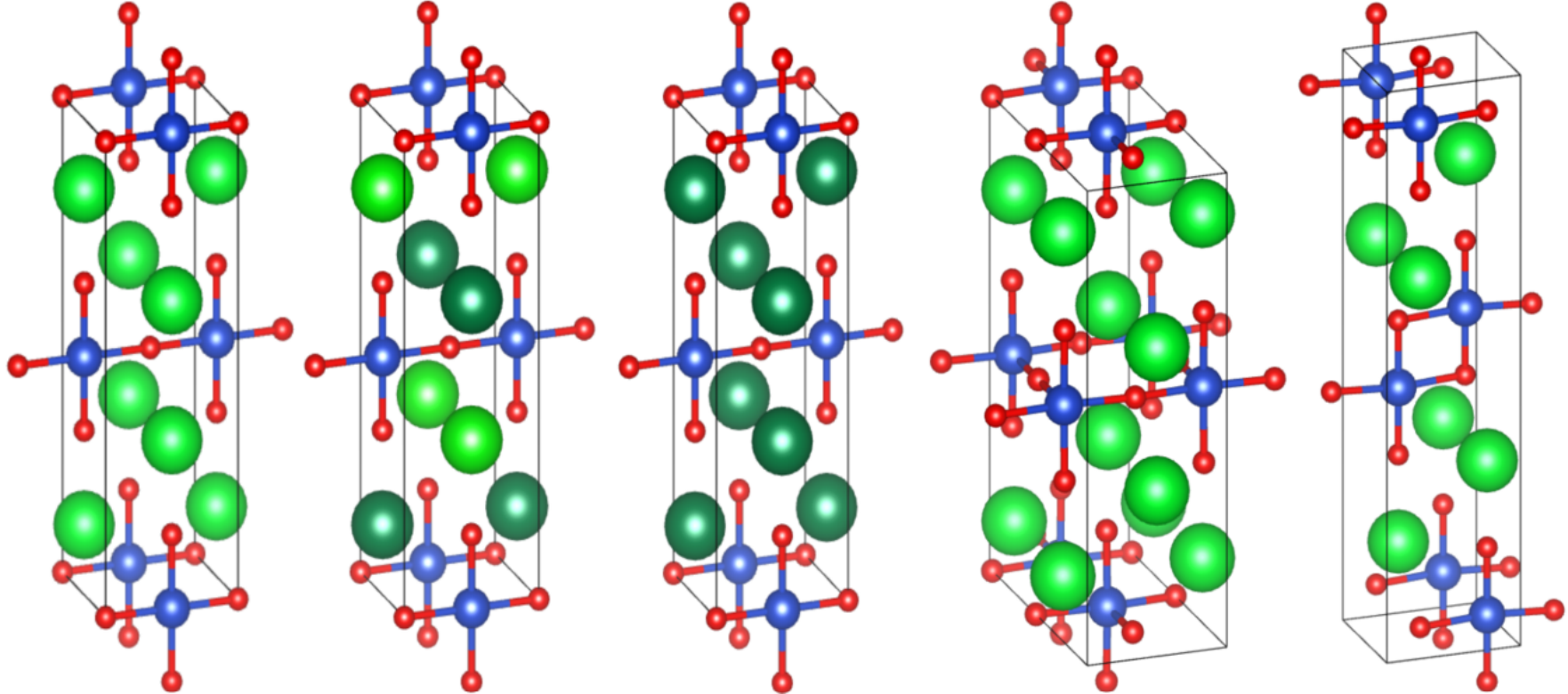}
};

\node[anchor=south west] at (14,0) {
  \includegraphics[width=0.06\textwidth]{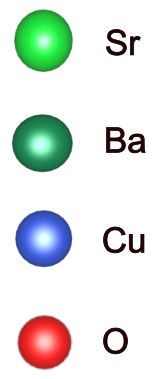}
};
\end{tikzpicture}
\caption{
Conventional unit cells of $\mathrm{Sr_2CuO_3}$ (first), $\mathrm{SrBaCuO_3}$ (second), 
$\mathrm{Ba_2CuO_3}$ (third), $\mathrm{Sr_2CuO_{3.5}}$ (fourth), and $\mathrm{SrCuO_2}$ (fifth). 
In the first three materials, the Cu-O chains run along the crystallographic $a$ direction. 
In $\mathrm{Sr_2CuO_{3.5}}$, the Cu-O network forms a ladder structure located in the $ac$ plane, whereas in $\mathrm{SrCuO_2}$ the ladder lies in the $ab$ plane.}
\label{fig:unitcells}
\end{figure*}

The present work focuses on including interchain and interladder interactions to achieve a more accurate description of strongly correlated quasi-1D materials. Our study focuses on the compounds \(\mathrm{Sr}_2\mathrm{CuO}_3\), \(\mathrm{SrBaCuO}_3\), and \(\mathrm{Ba}_2\mathrm{CuO}_3\), which are composed of infinite chains of corner-sharing \(\mathrm{CuO}_4\) plaquettes embedded within an otherwise 3D crystal lattice, as illustrated in Fig.~\ref{fig:unitcells}.
These chains are well isolated, enabling an effective 1D low-energy description. \(\text{Sr}_2\text{CuO}_{3}\) serves as an important reference system for understanding the mechanisms of superconductivity in cuprates through oxygen doping \cite{Liu2006}. Its antiferromagnetic spin correlations along the chains and the nature of its low-lying excitations are observables that can be directly accessed through TN calculations of the Hubbard model.

In addition, we consider the cuprate materials $\mathrm{Sr}_2\mathrm{CuO}_{3.5}$ and $\mathrm{SrCuO}_2$, for which two-leg ladder-based effective descriptions have been introduced as a conceptual modeling framework.
For these systems, we apply the MPS + MF framework developed by Bollmark \textit{et al.} \cite{Adrian2023}. This framework has previously been applied to attractive Hubbard chains to study the competition between charge-density wave (CDW) order and superconductivity (SC), as well as to investigate the emergence of transient non-equilibrium SC \cite{Adrian2023_2,Adrian2025}. However, it needs to be investigated to what extent this formalism is applicable to the materials under consideration here.

\section{\label{sec:level1}II. Methodology}

Our methodology consists of three steps. First, we construct effective multiband Hubbard models using the DFT + cRPA approach. Second, we incorporate the interchain or interladder interactions within a MF approximation. Third, we solve the resulting effective Hamiltonian using MPS. We begin by introducing the construction of the effective Hubbard model and then outline the theoretical frameworks underlying the MPS + MF approaches. Finally, we describe how TN techniques are employed to obtain the low-energy properties of the system. The first step was discussed in detail in our previous work, Ref.~\cite{Vrancken2025}.

\subsection{\label{sec:level2}A. Downfolding to a Hubbard model}

Consider a multiband Hubbard model Hamiltonian of the form:
\begin{align}
\hat{H}
&= - \sum_{i,j} \sum_{\alpha,\beta} \sum_{\sigma}
t_{i\alpha,j\beta}\,
\hat{c}_{i\alpha\sigma}^\dagger
\hat{c}_{j\beta\sigma} \nonumber \\
&\quad + \frac{1}{2}
\sum_{i,j,k,l}
\sum_{\alpha,\beta,\gamma,\delta}
\sum_{\sigma,\sigma'}
\hat{c}^\dagger_{i\alpha\sigma}
\hat{c}_{j\beta\sigma}
\,U_{i\alpha, j\beta, k\gamma, l\delta}\,
\hat{c}^\dagger_{k\gamma\sigma'}
\hat{c}_{l\delta\sigma'}.
\end{align}
Here, the indices \( i,j,k,l \) label the lattice sites, while the Greek indices
\( \alpha,\beta,\gamma,\delta \) denote the orbital degrees of freedom at each site.
The indices \( \sigma \) and \( \sigma' \) refer to the electron spin projections.
The hopping parameter \( t_{i\alpha,j\beta} \) describes the transfer of an electron
with spin \( \sigma \) from orbital \( \beta \) on site \( j \) to orbital
\( \alpha \) on site \( i \).
The interaction tensor \( U_{i\alpha, j\beta, k\gamma, l\delta} \) represents the electron-electron interaction between orbitals \( \alpha,\beta,\gamma,\delta \) on sites \( i,j,k,l \), respectively.
Finally, \( \hat{c}^\dagger_{i\alpha\sigma} \) and \( \hat{c}_{i\alpha\sigma} \) are the fermionic creation and annihilation operators for an electron with spin \( \sigma \) in orbital \( \alpha \) at lattice site \( i \).

The first step is to determine the parameters in the Hubbard Hamiltonian such that the resulting model accurately describes the low-energy subspace of the system.
Starting from a DFT calculation, the low-energy subspace is selected, and the interactions between subspaces are approximated using cRPA. 
Subsequently, maximally localized Wannier functions (MLWFs) are employed to define the low-energy orbitals \cite{Marzari2012}.

Generally, Wannier functions have the following form

\begin{equation}
    \phi_{\mathbf{R}\alpha}(\mathbf{r}) = \frac{V}{(2\pi)^3} \int_{\text{BZ}} d\mathbf{k} \, e^{-i\mathbf{k} \cdot \mathbf{R}} \sum_{\beta} T_{\alpha,\beta}(\mathbf{k}) \psi_{\beta,\mathbf{k}}(\mathbf{r}),
\end{equation}

where $\psi_{\beta\mathbf{k}}$ is a Bloch function, \( V \) is the volume of the unit cell, \( \mathbf{R} \) is the real-space lattice vector, \( \alpha \) is the band index, and the integral is performed over the first Brillouin zone (BZ). The unitary matrix \(T_{\alpha\beta}\) represents the gauge freedom in the choice of Wannier functions. Minimizing the sum of the quadratic spreads of the Wannier functions eliminates the non-uniqueness of the Wannier functions, resulting in a set of MLWFs.

Using these MLWFs, the tight-binding hopping parameters can be calculated as

\begin{equation}
t_{i\alpha,j\beta}= - \int d\mathbf{r}\,\phi_{i\alpha}^*(\mathbf{r})\,\hat{H}^{\mathrm{KS}}\,\phi_{j\beta}(\mathbf{r}) ,
\end{equation}

where $\hat{H}^{\text{KS}}$ is the Kohn-Sham Hamiltonian \cite{Hohenberg1964, Kohn1965}. Since $\hat{H}^{\text{KS}}$ contains the Hartree and exchange-correlation potentials constructed from the full electronic Hilbert space, its direct use in a low-energy effective model leads to a double-counting problem, as interaction effects within the low-energy subspace are already included. The constrained GW (cGW) approximation provides a systematic way to eliminate this double counting \cite{Hirayama2013}, its implementation is left for future work.
The Coulomb matrix elements \( U_{i\alpha, j\beta, k\gamma, l\delta} \) are integrals of the following form

\begin{align}
U_{i\alpha,\, j\beta,\, k\gamma,\, l\delta} (\omega)
&= \int \int d\mathbf{r}\, d\mathbf{r'}\,
\phi_{i\alpha}^*(\mathbf{r})\,
\phi_{j\beta}(\mathbf{r}) \nonumber \\
& \quad W_r(\mathbf{r}, \mathbf{r'}, \omega)\,
\phi_{k\gamma}^*(\mathbf{r'})\,
\phi_{l\delta}(\mathbf{r'}) .
\end{align}

The partially screened interaction \( W_r(\mathbf{r}, \mathbf{r'}, \omega) \) is computed using the cRPA, further details can be found in Ref.~\cite{Aryasetiawan2004, Aryasetiawan2006, Vrancken2025}. In the remainder of this work, we set \(\omega = 0\) and consider only the static limit. Although neglecting the frequency dependence can lead to non-negligible effects \cite{Hirayama2013, Scott2024}, we leave the inclusion of dynamical corrections for future work.

\subsection{\label{sec:level2}B. MPS + MF framework for chain systems}

When working with quasi-1D materials such as \(\text{Sr}_2\text{CuO}_3\), the interchain interactions can be incorporated through a MF coupling scheme. While such MF approaches have been widely employed in model studies \cite{Sandvik1999},  their application to \emph{ab initio}-derived descriptions of real materials remains largely unexplored.
In the following, we restrict ourselves to the one-band Hubbard model at half filling, which provides an effective low-energy description of the cuprate chains. 
Although the MF construction can, in principle, be generalized to an arbitrary 3D network of coupled chains, for the materials considered here, it is sufficient to retain only the coupling to the two nearest neighboring chains since hopping to more distant chains is negligible. These nearest neighboring chains are located along the crystallographic \(c\)-direction, as shown in Fig.~\ref{fig:unitcells}.

\subsubsection{\label{sec:level3}1. Replacing interchain Hubbard interactions with an effective spin model}

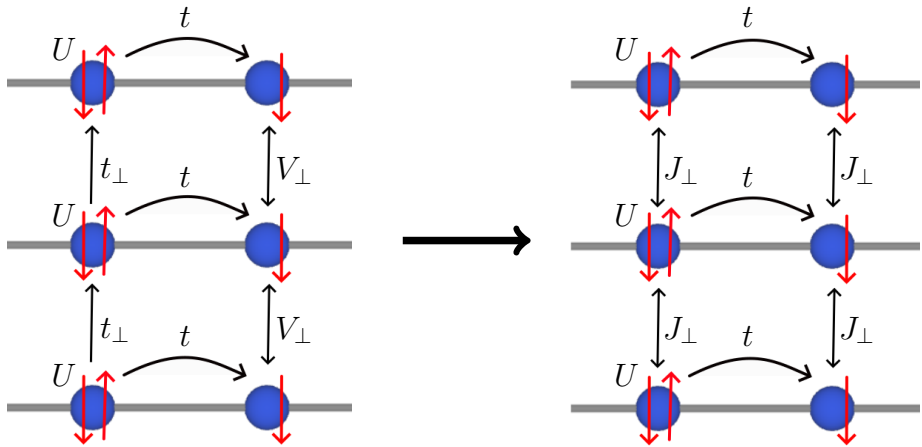
\begin{figure*}
\centering
\begin{tikzpicture}[scale=0.85, transform shape]
\node[anchor=south west] (image) at (-7.0,0.0) {
\includegraphics[width=0.3\linewidth]{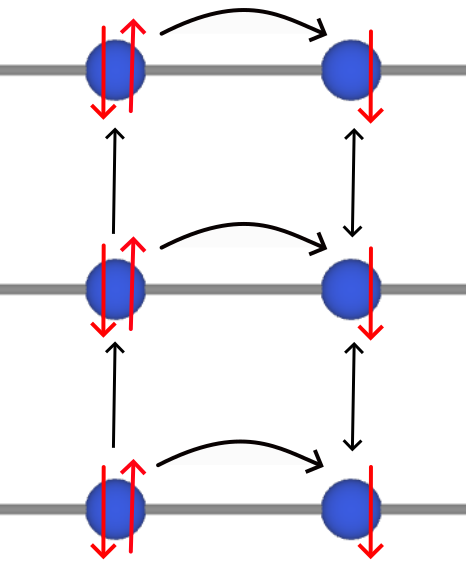}};
\node[anchor=south west] (image) at (1.8,0.0) {
\includegraphics[width=0.305\linewidth]{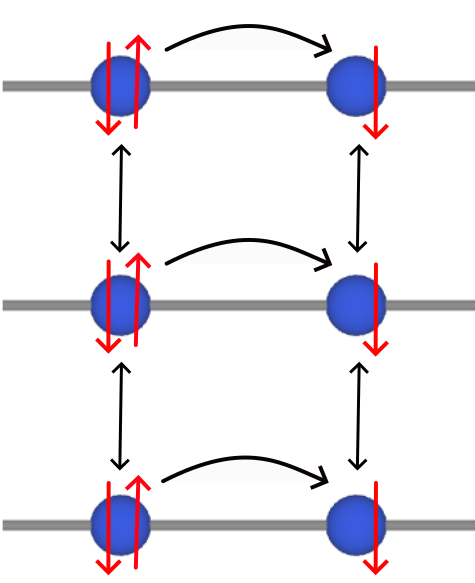}};
\draw[->, line width=1.0mm, black] (-0.7,3.4) -- (1.3,3.4);

\node at (-6.0,6.4) {\Large $U$};
\node at (-6.0,3.8) {\Large $U$};
\node at (-6.0,1.3) {\Large $U$};
\node at (-4.1,6.9) {\Large $t$};
\node at (-4.1,4.4) {\Large $t$};
\node at (-4.1,1.9) {\Large $t$};
\node at (-5.2,4.5) {\Large $t_\perp$};
\node at (-5.2,2.0) {\Large $t_\perp$};
\node at (-2.4,4.5) {\Large $V_\perp$};
\node at (-2.4,2.0) {\Large $V_\perp$};
\node at (2.8,6.4) {\Large $U$};
\node at (2.8,3.8) {\Large $U$};
\node at (2.8,1.3) {\Large $U$};
\node at (4.7,6.9) {\Large $t$};
\node at (4.7,4.4) {\Large $t$};
\node at (4.7,1.9) {\Large $t$};
\node at (3.65,4.5) {\Large $J_\perp$};
\node at (3.65,2.0) {\Large $J_\perp$};
\node at (6.4,4.5) {\Large $J_\perp$};
\node at (6.4,2.0) {\Large $J_\perp$};
\end{tikzpicture}

\caption{The interchain Hubbard interactions $t_\perp$ and $V_\perp$ are replaced by an effective Heisenberg exchange $J_\perp$. Here $t$ and $U$ denote the intrachain hopping and on-site Coulomb interaction, while $t_\perp$, $V_\perp$, and $J_\perp$ describe interchain hopping, interchain Coulomb interaction, and spin exchange, respectively.}
\label{firststepcoupling}
\end{figure*}

At half filling, when the interchain hopping \(t_\perp\) is much smaller than the onsite screened interaction \(U = U_{i,i,i,i}\), the interchain Hubbard interactions may be replaced by an effective spin-\(\frac{1}{2}\) Heisenberg description, as illustrated in Fig.~\ref{firststepcoupling}. The total Hamiltonian becomes
\begin{equation}
\hat{H}_{\text{total}} 
= \sum_{R_k} \hat{H}_{R_k} + \hat{H}_{\perp},
\end{equation}
where \(\hat{H}_{R_k}\) is the Hubbard Hamiltonian of chain \(R_k\). The interchain Hamiltonian is described by
\begin{equation}
\hat{H}_{\perp} 
= J_{\perp} \sum_{i,R_{k}} 
\vec{S}_{i,R_{k}}\cdot \vec{S}_{i,R_{k+1}}.
\end{equation}
where \( \vec{S}_{i,R_k} \) denotes the spin operator at site \(i\) of chain \(R_k\). The interchain antiferromagnetic exchange coupling $J_{\perp}$ can be derived from a Schrieffer-Wolff transformation to second and fourth order in \(t_\perp/U\) ~\cite{Anderson1950, Anderson1959, Macdonald1988}, yielding  
\begin{equation}
J_{\perp} 
= \frac{4 t_\perp^{2}}{U-V_\perp}
  - \frac{24 t_\perp^{4}}{(U-V_\perp)^{3}}
  + \mathcal{O}\!\left(\frac{t_\perp^{6}}{(U-V_\perp)^{5}}\right).
\label{eq:Jperp}
\end{equation}

with \(V_{\perp} = U_{i,i,j,j}\) the interchain screened Coulomb repulsion.

\subsubsection{\label{sec:level3}2. Mean-field decoupling of interchain interactions}

\begin{figure*}
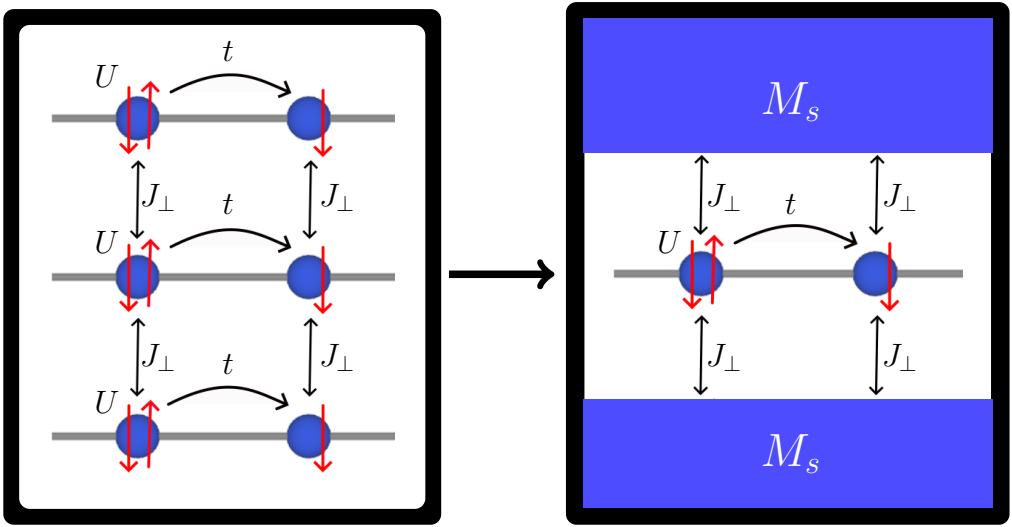

\centering
\begin{tikzpicture}[scale=0.85, transform shape]
\node[anchor=south west] (image) at (-7.0,-2.5) {
\includegraphics[width=0.3\linewidth]{Figures/3chainswithJ.png}};
\begin{scope}[on background layer]
\node[fill=black, rounded corners, fit=(image), inner sep=15pt] {};
\node[fill=white, rounded corners, fit=(image), inner sep=9pt] {};
\end{scope}
\node[anchor=south west] (image) at (1.8,-2.5) {
\includegraphics[width=0.305\linewidth]{Figures/3chainswithJ.png}};
\begin{scope}[on background layer]
\node[fill=black, rounded corners, fit=(image), inner sep=15pt] {};
\node[fill=white, rounded corners, fit=(image), inner sep=8pt] {};
\end{scope}
\begin{scope}
\node[
  fill=blue!70,
  minimum width=6.4cm,
  minimum height=2.1cm
] at (4.65,3.75) {};
\end{scope}
\begin{scope}
\node[
  fill=white,
  minimum width=2.1cm,
  minimum height=0.4cm
] at (4.65,-1.0) {};
\end{scope}
\begin{scope}
\node[
  fill=blue!70,
  minimum width=6.4cm,
  minimum height=1.7cm
] at (4.65,-2.0) {};
\end{scope}
\draw[->, line width=1.0mm, black] (-0.65, 0.8) -- (1.0, 0.8);
\node at (-6.0,3.9) {\Large $U$};
\node at (-6.0,1.3) {\Large $U$};
\node at (-6.0,-1.2) {\Large $U$};
\node at (-4.1,4.3) {\Large $t$};
\node at (-4.1,1.9) {\Large $t$};
\node at (-4.1,-0.6) {\Large $t$};
\node at (-5.2,2.0) {\Large $J_\perp$};
\node at (-5.2,-0.5) {\Large $J_\perp$};
\node at (-2.4,2.0) {\Large $J_\perp$};
\node at (-2.4,-0.5) {\Large $J_\perp$};

\node at (2.8,1.3) {\Large $U$};
\node at (4.7,1.9) {\Large $t$};
\node at (3.65,2.0) {\Large $J_\perp$};
\node at (3.65,-0.5) {\Large $J_\perp$};
\node at (6.4,2.0) {\Large $J_\perp$};
\node at (6.4,-0.5) {\Large $J_\perp$};

\node[text=white] at (4.7,3.5) {\huge $M_s$};
\node[text=white] at (4.7,-2.0) {\huge $M_s$};

\end{tikzpicture}
\caption{The nearest neighbor chains are replaced by an effective MF with the staggered magnetization $M_s$ as parameter. Here $t$ and $U$ denote the intrachain 
hopping and on-site Coulomb interaction, while $J_\perp$ describes spin exchange.}
\label{secondstepcoupling}
\end{figure*}

In the second step, the coupling to other chains is incorporated through a MF approximation. Each neighboring chain is treated independently and characterized by its staggered magnetization \(M_s\), which generates an effective external field acting on the chain under consideration, as illustrated in Fig.~\ref{secondstepcoupling}.

The staggered magnetization is defined as
\begin{equation}
M_s = \frac{1}{N} \sum_i (-1)^i \langle S_i^z \rangle,
\label{eq:staggered_mag}
\end{equation}
where \(N\) is the number of sites in a chain.

Within MF theory, we decouple the interchain coupling by assuming antiferromagnetic correlations between neighboring chains, corresponding to a N\'eel ordering pattern in the interchain direction. The interchain Hamiltonian then becomes~\cite{Sandvik1999}
\begin{align}
\hat{H}_{\perp}
&\approx J_{\perp}\sum_{i,R_k} S_{i,R_k}^{z}
\left( \langle S_{i,R_{k-1}}^{z} \rangle
     + \langle S_{i,R_{k+1}}^{z} \rangle \right) \nonumber \\
&= 2 J_{\perp} M_s \sum_{i} (-1)^i S_i^z .
\label{eq:interchain_H}
\end{align}

This follows from $\langle S_{i,R_k}^z \rangle = (-1)^{i+R_k} M_s$.

A challenge in our approach remains: we require $M_s$ to introduce interchain interactions, yet $M_s$ is not known \textit{a priori}. We will thus need a self-consistent loop to determine $M_s$ \cite{Sandvik1999}. We begin by computing the ground state using an initial guess for the interchain interaction (\( M_s \neq 0 \)), and then extract the corresponding \( M_s \) from this ground state. This updated \( M_s \) is used to redefine the external staggered MF, leading to a new ground state with a new value of \( M_s \). Repeating this process, \( M_s \) eventually converges to a stable value. After convergence, the resulting $M_s$ is analyzed qualitatively and compared across the quasi-1D materials to identify possible trends.

\vspace{1.9em}
\subsection{\label{sec:level2}C. MPS + MF framework for ladder systems}

In this subsection, we briefly outline the MPS + MF approach and the perturbative framework developed by Bollmark \textit{et al.}~\cite{Adrian2023}. 
The goal is to investigate whether this framework can be meaningfully applied to two-leg ladder-based effective models of the materials studied here.

To describe our materials beyond isolated ladders, we consider a lattice of coupled ladders that allows for interladder tunneling and correlated motion of electrons across the array. The total Hamiltonian is thus given by a sum over ladder contributions and their couplings:

\begin{equation}
H_{\text{tot}} = \sum_{k,l} H_{\text{ladder}}(R_{k,l}) + H_\perp
\label{Htot}
\end{equation}

\begin{figure}
\centering
\begin{tikzpicture}
\node[anchor=south west, inner sep=0] (img) at (0,0)
{\includegraphics[width=\linewidth]{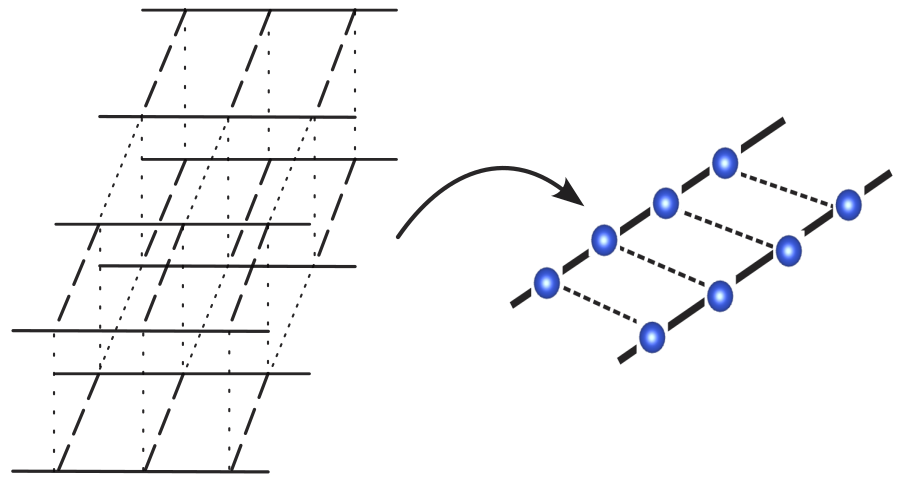}};
\begin{scope}[x={(img.south east)}, y={(img.north west)}]
\node[anchor=east, fill=white, inner sep=2pt] at (0.125,0.65) {$t_\perp$};
\node[anchor=east, fill=white, inner sep=2pt] at (0.055,0.15) {$t_{ab}$};
\node[anchor=east, fill=white, inner sep=2pt] at (0.29,1.03) {$t_{leg}$};
\node[anchor=east, fill=white, inner sep=2pt] at (0.175,0.9) {$t_{rung}$};
\draw[<->, ultra thick, red]
(0.70,0.20) -- (0.70,0.05)
node[midway, anchor=east, fill=white, inner sep=2pt]
{\textcolor{red}{$\alpha_{i,i',j,j'}$}};
\draw[<->, ultra thick, green!60!black]
(0.825,0.35) .. controls (0.88,0.22) and (0.91,0.22) .. (0.94,0.5)
node[pos=0.5, sloped, below, fill=white, inner sep=2pt]
{\textcolor{green!60!black}{$\beta_{i,i',j,j'}$}};
\node[anchor=east, fill=white, inner sep=2pt, rotate=40] at (0.78,0.73) {$t_{leg}$};
\node[anchor=east, fill=white, inner sep=2pt, rotate=-25] at (0.69,0.26) {$t_{rung}$};
\end{scope}
\end{tikzpicture}
\caption{
Schematic illustration of the MPS + MF framework, depicting the mapping from a lattice of coupled ladders to an effective single-ladder description.
All intra- and interladder hopping processes included in the model are indicated.}
\label{fig:ladder_framework}
\end{figure}
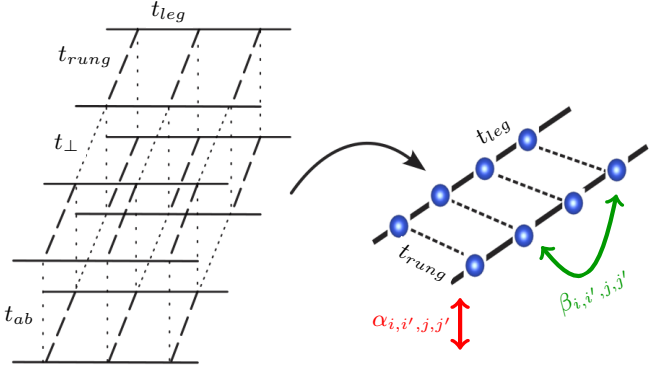

Here, $H_{\text{ladder}}$ represents the local dynamics within a single ladder, while $H_\perp$ encodes tunneling processes between neighboring ladders in the array. For the sake of clarity, we repeat the most important concepts and formulas of the used MPS + MF approach. For more details we refer to the work of Bollmark \textit{et al.}~\cite{Adrian2023}. A schematic illustration of this framework showing how an array of coupled ladders is mapped onto an effective single-ladder description and indicating the relevant intra- and interladder hopping processes, is provided in Fig.~\ref{fig:ladder_framework}.

\begin{widetext}
\begin{align}
H_{\text{ladder}}(R_{k,l})
= -t_{\mathrm{leg}} \sum_{j=0}^{1} \sum_{i,\sigma}
\left( c^{\dagger}_{i+1,j,\sigma}(R_{k,l})\, c_{i,j,\sigma}(R_{k,l})
+ \text{h.c.} \right)
- t_{\mathrm{rung}} \sum_{i,\sigma}
\left( c^{\dagger}_{i,1,\sigma}(R_{k,l})\, c_{i,0,\sigma}(R_{k,l})
+ \text{h.c.} \right) \nonumber \\
+ U_{i,i,i,i} \sum_{i,j}
n_{i,j,\uparrow}(R_{k,l})\, n_{i,j,\downarrow}(R_{k,l})
+ U_{i,i,i+1,i+1} \sum_{i} 
\Big[\sum_{j=0}^{1} n_{i,j}(R_{k,l})\, n_{i+1,j}(R_{k,l})
+ n_{i,0}(R_{k,l})\, n_{i,1}(R_{k,l})
\Big].
\end{align}

\begin{align}
H_{\perp} 
&= - t_{\perp} \sum_{i} \sum_{k,l} \sum_{\sigma}
\left[ c^{\dagger}_{i,1,\sigma}(R_{k,l})\, c_{i,0,\sigma}(R_{k+1,l})
+ c^{\dagger}_{i,0,\sigma}(R_{k+1,l})\, c_{i,1,\sigma}(R_{k,l}) \right] \nonumber \\
&\quad - t_{ab} \sum_{i} \sum_{j=0}^{1} \sum_{k,l} \sum_{\sigma}
\left[ c^{\dagger}_{i,j,\sigma}(R_{k,l})\, c_{i,j,\sigma}(R_{k,l+1})
+ c^{\dagger}_{i,j,\sigma}(R_{k,l+1})\, c_{i,j,\sigma}(R_{k,l}) \right].
\end{align}
\end{widetext}
The index \( i \) labels the sites along the legs of the ladder, while \( j = 0, 1 \) labels the two sites on each rung. Here \( t_{\mathrm{leg}} \) and \( t_{\mathrm{rung}} \) are the intraladder hopping parameters along the legs and across the rungs, respectively.

When the interladder hopping $t_{\perp}$ and $t_{ab}$ are small compared to the intrinsic energy scales of an isolated ladder, namely the pairing energy $\Delta E_p$ and twice the spin gap $2\Delta E_s$, the effects of interladder coupling can be treated perturbatively. The pairing energy $\Delta E_p$ and the spin gap $\Delta E_s$ are defined as

\begin{equation}
\label{Ep}
\Delta E_p = 2E(N + 1, \tfrac{1}{2}) - E(N, 0) - E(N + 2, 0)
\end{equation}

\begin{equation}
\label{Es}
\Delta E_s
= E(N, 1) - E(N, 0),
\end{equation}

with $E(N,S)$ denoting the ground-state energy of a single ladder with particle number $N$ and total spin $S$. The pairing energy $\Delta E_p$ quantifies the effective binding energy of two particles within a ladder, while the spin gap $\Delta E_s$ measures the energy cost of creating a spin-triplet excitation.
In the small $t_{\perp}/\Delta E_p$ regime, an effective low-energy Hamiltonian can be derived from a Schrieffer-Wolff transformation \cite{Adrian2023}. The resulting description captures the leading-order processes that mediate interladder coupling through virtual excitations of bound pairs.

\begin{equation}
H_{\mathrm{eff}} = H_{\text{ladder}} - \frac{1}{\Delta E_p} \, P_0 H_\perp^2 P_0 + \mathcal{O}(H_\perp^4).
\end{equation}
where $P_0$ is the projector onto the low-energy manifold of decoupled ladders, corresponding to the ground-state sector of each individual ladder.

We now assume that all ladders are identical and that the MF is real-valued \cite{Adrian2023}. Furthermore, we neglect expectation values of operators that create or annihilate particles on different ladders, reflecting the assumption that the constituents of a bound pair cannot reside on separate ladders.

Within the MPS + MF framework, the resulting effective Hamiltonian for a single ladder now takes the form \cite{Adrian2023}: 

\begin{equation}
H_{\text{eff,MF}} = H_{\text{ladder}} - H_{\text{pair,MF}} + H_{\text{exc,MF}}
\end{equation}

with

\begin{equation}
H_{\text{pair,MF}} = 
\sum_{i,i',j,j'} \alpha_{i,i',j,j'}
\big( c^{\dagger}_{i,j,\uparrow} c^{\dagger}_{i',j',\downarrow} 
+ c_{i',j',\downarrow} c_{i,j,\uparrow} \big),
\end{equation}

\begin{equation}
H_{\text{exc,MF}} = \sum_{i,i',j,j',\sigma} 
\beta_{i,i',j,j',\sigma} 
c^{\dagger}_{i,j,\sigma} c_{i',j',\sigma},
\end{equation}

The pairing amplitudes, describing pair tunneling into or out of the ladder, are given by \cite{Adrian2023}:

\begin{equation}
\alpha_{i,i',0,0} = 
\frac{2}{\Delta E_p}
\Big( 
t_{\perp}^2\langle c_{i',1,\downarrow} c_{i,1,\uparrow} \rangle 
+ 2 t_{ab}^2 \langle c_{i',0,\downarrow} c_{i,0,\uparrow} \rangle 
\Big),
\end{equation}

\begin{equation}
\alpha_{i,i',1,1} = 
\frac{2}{\Delta E_p}
\Big( t_{\perp}^2
\langle c_{i',0,\downarrow} c_{i,0,\uparrow} \rangle 
+ 2t_{ab}^2 \langle c_{i',1,\downarrow} c_{i,1,\uparrow} \rangle 
\Big),
\end{equation}

\begin{equation}
\alpha_{i,i',1,0} = 
\frac{4 t_{ab}^2}{\Delta E_p}
\langle c_{i',0,\downarrow} c_{i,1,\uparrow} \rangle,
\end{equation}

\begin{equation}
\alpha_{i,i',0,1} = 
\frac{4 t_{ab}^2}{\Delta E_p}
\langle c_{i',1,\downarrow} c_{i,0,\uparrow} \rangle,
\end{equation}

whereas the exchange terms are given by \cite{Adrian2023}:

\begin{equation}
\beta_{i,i',0,0,\sigma} = 
\frac{2}{\Delta E_p}
\Big( t_{\perp}^2
\langle c^{\dagger}_{i,1,\sigma} c_{i',1,\sigma} \rangle 
+ 2t_{ab}^2 \langle c^{\dagger}_{i,0,\sigma} c_{i',0,\sigma} \rangle 
\Big).
\end{equation}

\begin{equation}
\beta_{i,i',1,1,\sigma} = 
\frac{2}{\Delta E_p}
\Big( t_{\perp}^2
\langle c^{\dagger}_{i,0,\sigma} c_{i',0,\sigma} \rangle 
+ 2t_{ab}^2 \langle c^{\dagger}_{i,1,\sigma} c_{i',1,\sigma} \rangle 
\Big),
\end{equation}

\begin{equation}
\beta_{i,i',1,0,\sigma} = 
\frac{4 t_{ab}^2}{\Delta E_p}
\langle c^{\dagger}_{i,1,\sigma} c_{i',0,\sigma} \rangle,
\end{equation}

\begin{equation}
\beta_{i,i',0,1,\sigma} = 
\frac{4 t_{ab}^2}{\Delta E_p}
\langle c^{\dagger}_{i,0,\sigma} c_{i',1,\sigma} \rangle.
\end{equation}

The MF parameters $\alpha$ and $\beta$ are obtained self-consistently: starting from an initial guess, we compute the ground state of $H_{\text{eff,MF}}$, update the expectation values, and repeat until convergence. All simulations were carried out using our in-house Julia code \texttt{HubbardTN.jl}, which implements the approach described above and is publicly available~\cite{HubbardTNrepo}.

\subsection{\label{sec:level2}D. Solving the Hubbard model}

Having constructed the Hubbard model, we now introduce the matrix product state framework and the algorithms used to compute its ground state. 

Consider a lattice of \( N \) physical spins. 
Although we naturally consider a one-dimensional lattice, the following also applies to lattices of arbitrary dimension where sites have been numbered. However, MPS representations of states on higher-dimensional lattices become impractical for numerical computations. 
Every site has a local Hilbert space \( \mathcal{H}_i = \mathbb{C}^d \) of dimension \( d \), this is called the physical dimension. 
For a Hubbard model, the local physical dimension is $d = 4$, corresponding to the states $\{ |0\rangle, |\uparrow\rangle, |\downarrow\rangle, |\uparrow\downarrow\rangle \}$. 
The total Hilbert space of the lattice is thus \( \mathcal{H} = \bigotimes_{i=1}^{N} \mathcal{H}_i = (\mathbb{C}^d)^{\otimes N} \). 
A general quantum state \( |\psi\rangle \) in this Hilbert space can be expressed using \( d^N \) complex coefficients \( C_{s_1,\dots,s_N} \in \mathbb{C} \), where \( s_i \in \{0, \dots, d-1\} \) \cite{Schollwock2011}:

\begin{equation}
\ket{\psi} = \sum_{s_1,\dots,s_N} C_{s_1,\dots,s_N} |s_1, \dots, s_N \rangle,
\end{equation}

The tensor $C_{s_1,\dots,s_N}$ can be factorized into smaller tensors that retain the essential structure of the quantum state while reducing storage and computational complexity. By applying multiple singular value decompositions (SVD), the tensor can be rewritten as a product of matrices \cite{Schollwock2011}, hence the name MPS.

\begin{equation}
\ket{\psi} = \sum_{s_1,\dots,s_N}
A^{[1]s_1} A^{[2]s_2} \dots A^{[N]s_N}
|s_1, \dots, s_N \rangle.
\end{equation}

Each site $i$ is associated with its own tensor $A^{[i]}$, whose physical index $s_i$ selects one of $d$ matrices of dimension $D_i \times D_{i+1}$.
Open boundary conditions imply $D_1 = D_{N+1} = 1$. 
The bond dimension $D$ controls the amount of entanglement captured by the MPS, increasing $D$ improves the accuracy of the MPS and becomes exact in the limit $D \to \infty$.
In the thermodynamic limit \( N \to \infty \), it becomes natural to consider translationally invariant quantum states. A uniform MPS (uMPS) \cite{Zauner2018} is a special MPS ansatz in which a finite set of matrices $\{A^{[1]}, \dots, A^{[\ell]}\}$ defining a unit cell of length $\ell$ is repeated periodically along the lattice.

\begin{align}
\ket{\psi} &= \sum_{\{s\}}\mathbf{v}_L^\dagger\left(\prod_{l}A^{[l] s}\right)\mathbf{v}_R\ket{s} \nonumber \\
&= \text{\ldots
\begin{tikzpicture}[baseline=-0.5ex]
\node[draw, rounded corners] (A) at (0,0) {$A^{[1] s_1}$};
\node[draw, rounded corners] (B) at (1.5,0) {$A^{[2] s_2}$};
\node at (2.7,0) {$\ldots$};
\node[draw, rounded corners] (C) at (3.9,0) {$A^{[l] s_l}$};
\draw (A) -- (B);
\draw (B) -- (2.4,0);
\draw (3.0,0) -- (C);
\draw (A) -- ++(-0.85, 0.0);
\draw (A) -- ++(0.0, -0.6);
\draw (B) -- ++(0.0, -0.6);
\draw (C) -- ++(0.0, -0.6);
\draw (C) -- ++(0.85, 0.0);
\end{tikzpicture} \, \ldots
}
\end{align}

The column vectors $v_L$ and $v_R$ encode the boundary conditions of the MPS. In the thermodynamic limit, these vectors carry no physical significance. We also show the Penrose graphical notation of the uMPS. In this notation, shapes represent tensors and lines (legs) correspond to their indices. Legs pointing downward are associated with the physical Hilbert space of dimension $d$, while horizontal legs represent the virtual spaces of dimension $D$. Connected legs indicate the contraction of the corresponding indices. 

Since its introduction by Steven White in 1992, the DMRG algorithm has become one of the most powerful numerical methods for studying one-dimensional quantum lattice systems \cite{White1992,White1993}. 
In this work, we find the lowest-energy matrix product state wavefunction of a Hamiltonian using infinite-system DMRG (iDMRG), which directly targets the thermodynamic limit.
The algorithm variationally optimizes a translationally invariant matrix product state by iteratively growing the unit cell and projecting it onto an effective low-dimensional Hilbert space, where the truncation is controlled by an SVD of the reduced density matrix.

An alternative matrix product state optimization scheme employed in this work is the variational uniform matrix product state (VUMPS) \cite{Zauner2018} algorithm, which is formulated within the tangent-space framework of uniform MPS.
Compared to iDMRG, VUMPS often exhibits faster and more stable convergence, particularly for systems with longer-range interactions \cite{Zauner2018}.
Accordingly, we employ both iDMRG and VUMPS to efficiently and reliably determine ground-state properties.

\section{\label{sec:level1}III. Computational Details}

\subsection{\label{sec:level2}A. Structural Relaxation}

All DFT calculations, including full structural optimizations and subsequent electronic-structure post-processing, were performed using the Vienna \textit{ab initio} Simulation Package (VASP) \cite{Kresse1993,Kresse1996,Hafner2008}. Exchange-correlation effects were treated within the generalized gradient approximation (GGA) using the Perdew-Burke-Ernzerhof (PBE) functional \cite{Perdew1996}. The interaction between valence electrons and ionic cores was described using the projector augmented-wave (PAW) method \cite{Blochl1994}. The following PAW potentials were used: \texttt{Sr\_sv} (4s, 4p, 5s; 10 valence electrons), \texttt{Ba\_sv} (5s, 5p, 6s; 10 valence electrons),  \texttt{Cu} (3d, 4s, 4p; 11 valence electrons), and \texttt{O} (2s, 2p; 6 valence electrons). Long-range van der Waals interactions were accounted for using the DFT-D3 dispersion correction with Becke-Johnson damping \cite{Grimme2011}.

The crystal structures of Sr$_2$CuO$_3$, Sr$_2$CuO$_{3.5}$, SrCuO$_2$, SrBaCuO$_3$, 
and Ba$_2$CuO$_3$ were fully optimized within VASP by relaxing both lattice parameters and internal atomic coordinates. Structural optimizations were performed using a $2 \times 1 \times 2$ supercell constructed from the conventional unit cell shown in Fig.~\ref{fig:unitcells}. The same computational parameters were used for all compounds, with Brillouin-zone integrations performed on a $7 \times 5 \times 7$ $k$-point mesh and a plane-wave energy cutoff of 600~eV.
Electronic self-consistency was achieved with a total energy convergence criterion of $10^{-8}$~eV, while ionic relaxation proceeded until the change in total energy between successive ionic steps was less than $10^{-7}$~eV. Gaussian smearing was applied to the electronic occupations during the structural optimization.

\subsection{\label{sec:level2}B. Downfolding}

Downfolding calculations for all materials were performed using the PBE functional in VASP on a $7 \times 3 \times 5$ $k$-point mesh with a plane wave cutoff of 600 eV. Screened Coulomb interaction parameters were computed using the cRPA as implemented in VASP. Since VASP evaluates interaction parameters only within the unit cell, supercells were constructed to calculate the nearest-neighbor interactions.

The MLWFs were constructed using the Wannier90 code \cite{Arash2008} interfaced with VASP through the VASP2Wannier90 workflow \cite{Hafner2008, Arash2008}.
For systems with entangled bands, the projector method was used within the cRPA 
formalism to define the correlated subspace. Visualization of the relaxed crystal structures and Wannier function isosurfaces was performed using the VESTA software \cite{Momma2008}.

\subsection{\label{sec:level2}C. Tensor Network Calculations}

To perform the calculations for the Hubbard model, we employed the open-source Julia packages \texttt{TensorKit.jl} \cite{TensorKit}, \texttt{MPSKit.jl} \cite{MPSKit}, and \texttt{MPSKitModels.jl}. These libraries provide efficient and flexible implementations of tensor-network algorithms, with built-in support for symmetries and fermionic degrees of freedom. All simulations were carried out using our in-house Julia code \texttt{HubbardTN.jl} and is publicly available \cite{HubbardTNrepo}.

In the MPS formulation of the Hubbard model, we introduce a filling
\begin{equation}
f = \frac{N_e}{N}, \qquad 0 \le f \le 2,
\end{equation}
defined as the number of electrons per site. Here, $N$ denotes the number of physical lattice sites in the material unit cell. The lattice is represented as an infinite one-dimensional chain, where the sites are arranged in a zigzag geometry, forming a strip of finite width. For a Hubbard model with \(B\) bands, this strip has a width of \(B\) sites, with each orbital treated as an individual site within the unit cell.
For the chain materials, we used a one-band Hubbard model ($B=1$), while for the two-leg ladder models, we used a two-band Hubbard model ($B=2$). In the ladder model, the rung index $j=0,1$ is treated as an effective band index, 
so that the two legs correspond to two bands.
To ensure the injectivity of the MPS, the MPS unit cell contains \(BN\) sites when \(N_e\) is even and \(2BN\) sites when \(N_e\) is odd \cite{Zauner2018}. In all calculations, a filling of \(f = 1\) was used.

For the chain materials, all calculations were performed using the \(U(1)\otimes U(1)\otimes f\mathbb{Z}_2\) symmetry, where the first \(U(1)\) enforces charge conservation, the second \(U(1)\) conserves the spin projection along a fixed axis and thus distinguishes spin-up and spin-down states, and \(f\mathbb{Z}_2\) accounts for the fermionic anticommutation relations. For the ladder materials, we employed a reduced symmetry structure consisting of \(f\mathbb{Z}_2\) for fermionic statistics, a single \(U(1)\) symmetry for spin conservation, and a trivial symmetry in the particle-number sector.

Once the Hubbard Hamiltonian is constructed, the ground state is obtained using the iDMRG2 algorithm \cite{White1992, White1993, Schollwock2011}, during which the bond dimension is determined dynamically. The resulting state is then further refined using several VUMPS iterations to ensure convergence. A gradient tolerance of $10^{-8}$ was used throughout.

\section{\label{sec:level1}IV. Results and discussion}

In this section, we first present the DFT + cRPA downfolding results for all materials considered in this work. We then present the MPS + MF results for the chain materials \(\mathrm{Sr}_2\mathrm{CuO}_3\), \(\mathrm{SrBaCuO}_3\), and \(\mathrm{Ba}_2\mathrm{CuO}_3\), where interchain couplings are included via a self-consistent staggered mean field. Finally, we assess the applicability of the ladder-based MPS + MF framework to \(\mathrm{Sr}_2\mathrm{CuO}_{3.5}\) and \(\mathrm{SrCuO}_2\).

\subsection{\label{sec:level2}A. Downfolding}

\begin{figure}[h]
\centering
\includegraphics[width=0.99\linewidth]{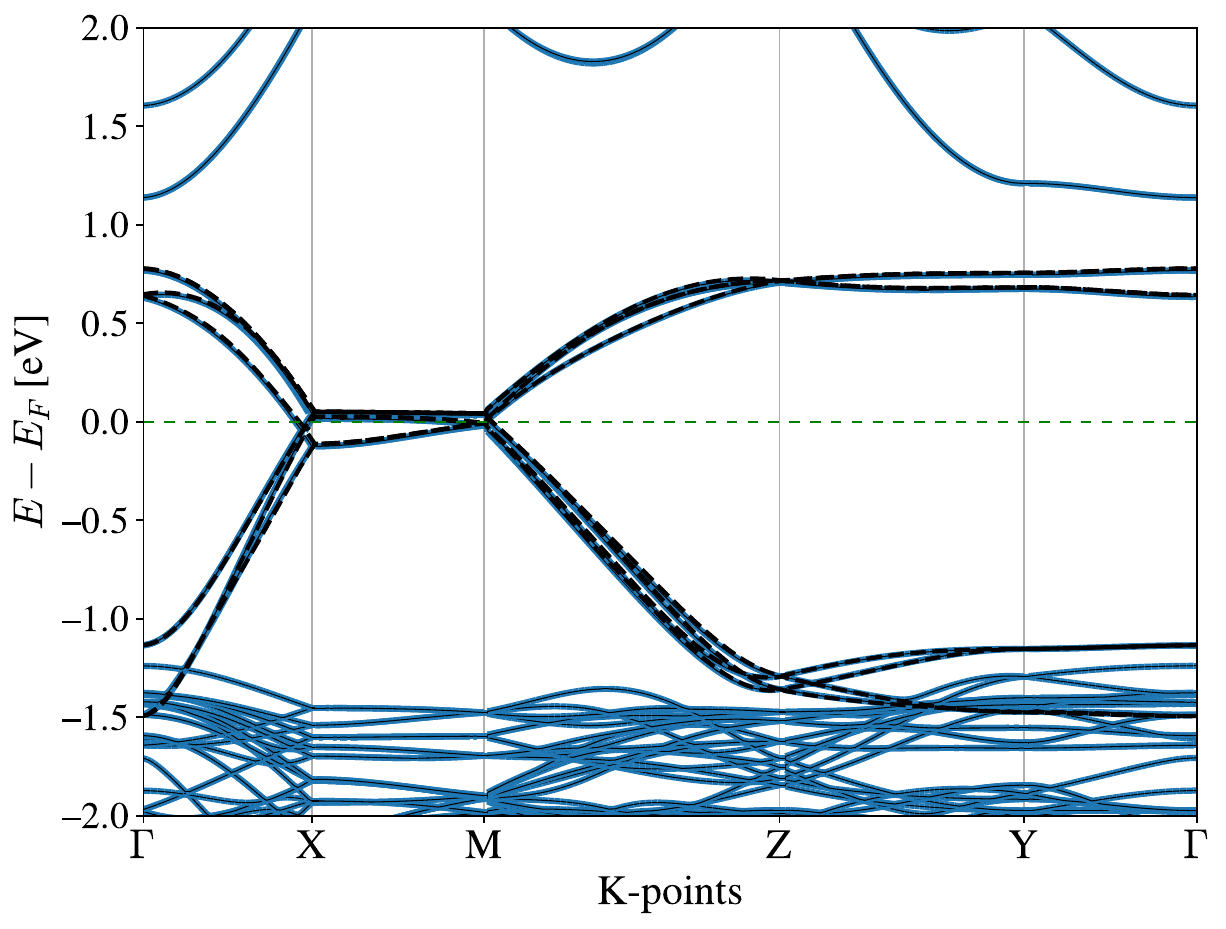}
\caption{The electronic band structure of \(\mathrm{Sr_2CuO_3}\), relative to the Fermi energy \(E_F\). The striped black lines represent the Wannier interpolated bands. An orthorhombic supercell containing eight formula units has been used. The high-symmetry points of the Brillouin zone are labeled as \(\Gamma=(0,0,0)\), \(X=(0.5,0,0)\), \(M=(0.5,0,0.5)\), \(Z=(0,0,0.5)\) and \(Y=(0,0.5,0)\).}
\label{fig:sr2cuo3-band}
\end{figure}

Starting from a KS-DFT calculation, a low-energy subspace that captures only the essential degrees of freedom associated with the strongly correlated electrons can be selected. Looking at the band structure of \(\text{Sr}_2\text{CuO}_3\) in Fig.~\ref{fig:sr2cuo3-band}, we can observe that only a few bands cross the Fermi energy. These upper Hubbard bands are primarily composed of Cu \(3d_{x^2 - y^2}\) orbitals \cite{Neudert2000}. (See Supplemental Information for orbital projections.)

\begin{figure*}
\centering
\begin{minipage}{0.48\textwidth}
\begin{tikzpicture}
\node[anchor=south west] (image) at (0.5,0) {
\includegraphics[width=0.5\textwidth,height=2.5cm,keepaspectratio,angle=0,scale=3.0]{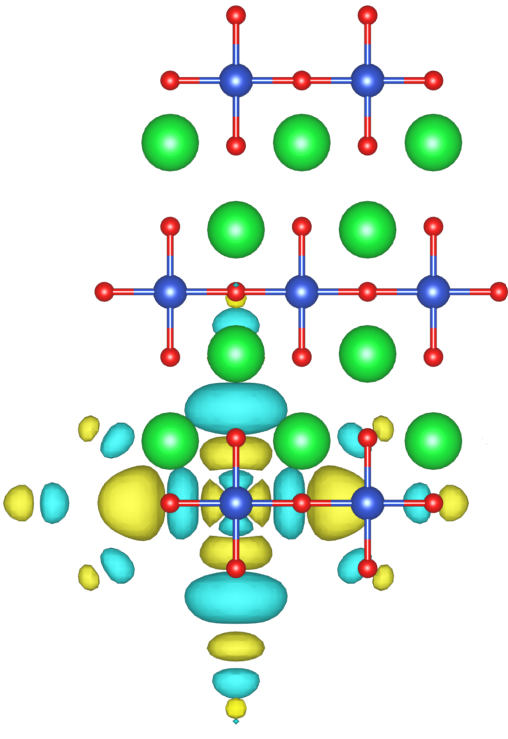}};
\draw[->, line width=0.7mm, black] (2.5, 4.9) to[bend left] node[midway, above = 0.93mm, font=\Large] {$t_2$} (4.8, 4.9);
\draw[->, line width=0.7mm, black] (3.2, 2.3) to[bend right] node[midway, above=-6mm, font=\Large] {$t$} (4.1, 2.3);
\draw[->, line width=0.7mm, black] (2.95, 2.7) -- (2.5, 4.4) node[midway, left, xshift=-5pt, yshift=4pt, font=\Large] {$t_{ab}$};
\end{tikzpicture}
\end{minipage}
\hfill
\begin{minipage}{0.48\textwidth}
\begin{tikzpicture}
\node[anchor=south west] (image) at (0.5,0) {
\includegraphics[width=0.5\textwidth,height=2.5cm,keepaspectratio,angle=0,scale=3.0]{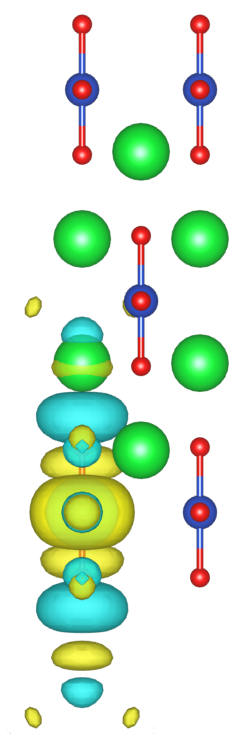}
\hspace{1.0cm}
\includegraphics[width=0.12\textwidth]{Materials/Legend.png}};
\draw[->, line width=0.7mm, black] (1.7, 2.6) to[bend left] node[midway, above = -7.0mm, xshift=5pt, font=\Large] {$t_{\perp}$} (2.5, 2.6);
\draw[->, line width=0.7mm, black] (1.45, 2.7) -- (1.45, 6.4) node[midway, left,xshift=15pt, font=\Large] {$t_b$};
\end{tikzpicture}
\end{minipage}
\caption[Schematic illustration of key hopping parameters.]{The MLWF isosurfaces of \( \text{Sr}_2\text{CuO}_3 \), illustrating key hopping parameters \(t, t_2, t_{\perp}, t_{ab}\) and \(t_b\).}
\label{fig:hoppingparameters}
\end{figure*}

A real-space visualization of the MLWF isosurface for \(\mathrm{Sr_2CuO_3}\), together with the hopping parameters, is shown in Fig.~\ref{fig:hoppingparameters}. The derived hopping and interaction parameters for all materials are summarized in Tab.~\ref{tab:hubbard_parameters} and Tab.~\ref{tab:hubbard_parameters2}. The interaction terms are defined in terms of the screened four-index Coulomb matrix elements
\(
U_{i,j,k,l}
\),
expressed in the basis of MLWFs. The onsite Coulomb repulsion is given by
\(
U = U_{i,i,i,i}
\),
while the nearest-neighbor Coulomb repulsion between sites \(i\) and \(j\) is
\(
V = U_{i,i,j,j}
\).
The corresponding exchange interaction is defined as
\(
J = U_{i,j,j,i} = U_{i,j,i,j}
\),
describing spin-flip and pair-hopping processes between nearest-neighbor sites.

\begin{table}[h]
\caption{Hopping and interaction parameters of the effective one-band Hubbard model for the quasi-1D cuprates studied in this work (all energies in eV).}
\label{tab:hubbard_parameters}
\begin{ruledtabular}
\begin{tabular}{lccc}
 & Sr$_2$CuO$_3$ & SrBaCuO$_3$ & Ba$_2$CuO$_3$ \\
\colrule
$t$        & 0.486 & 0.490 & 0.491 \\
$t_2$      & 0.077 & 0.082 & 0.077 \\
$t_3$      & 0.018 & 0.019 & 0.018 \\
$t_{\perp}$& 0.033 & 0.024 & 0.026 \\
$t_{ab}$   & 0.0068 & -0.0017 & -0.011 \\
$t_{b}$    & 0.0014 & 0.0033 & 0.0013 \\
$U$        & 3.411 & 3.238 & 2.980 \\
$V$        & 1.042 & 0.920 & 0.916 \\
$J$        & 0.033 & 0.031 & 0.038 \\
$V_{\perp}$& 0.799 & 0.666 & 0.649 \\
\colrule
$J_{\perp}$& 0.00167 & 0.000895 & 0.00116 \\
\end{tabular}
\end{ruledtabular}
\end{table}

\begin{table}
\caption{Hopping and interaction parameters of the effective two-band Hubbard model for the two-leg ladder-based models studied in this work (all energies in eV).}
\label{tab:hubbard_parameters2}
\begin{ruledtabular}
\begin{tabular}{lcc}
 & Sr$_2$CuO$_{3.5}$ & SrCuO$_2$ \\
\colrule
$t_{\mathrm{leg}}$  & 0.400 & 0.484 \\
$t_{\mathrm{rung}}$ & 0.066 & 0.00040\\
$t_{\perp}$ & 0.046 & 0.025 \\
$t_{ab}$    & 0.024 & 0.069 \\
$U$         & 1.123 & 3.495 \\
$V_{\mathrm{leg}}$  & 0.244 & 0.916 \\
$V_{\mathrm{rung}}$ & 0.119 & 0.239 \\
\end{tabular}
\end{ruledtabular}
\end{table}

\subsection{\label{sec:level2}B. MPS + MF framework for chains}

Using the Hubbard parameters obtained from the DFT + cRPA downfolding, we construct and solve an effective one-band Hubbard model to compute the ground state and the staggered magnetization of the quasi-1D materials. All intrachain hopping and interaction parameters listed in Tab.~\ref{tab:hubbard_parameters} are taken as direct input and fully define the Hamiltonian implemented in our MPS/iDMRG code. The interchain Hubbard interactions are incorporated separately through the MF treatment discussed above.

\begin{figure}
\centering
\includegraphics[width=0.95\linewidth]{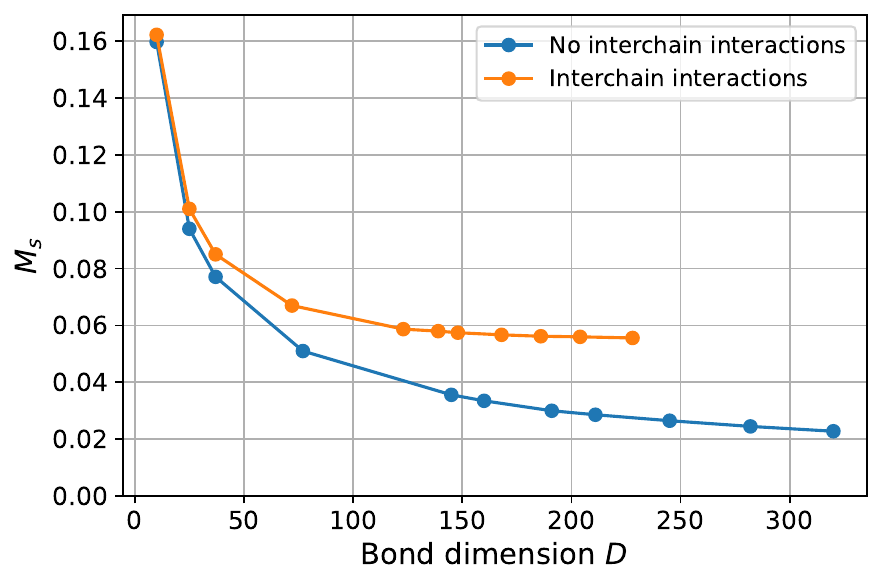}
\caption{The staggered magnetization \(M_s\) as a function of the bond dimension \(D\) for \(\mathrm{Sr_2CuO_{3}}\).}
\label{fig:Ms_vs_D-Sr2CuO3}
\end{figure}

In Fig.~\ref{fig:Ms_vs_D-Sr2CuO3}, we show the staggered magnetization \(M_s\), defined in Eq.~\ref{eq:staggered_mag}, as a function of the bond dimension \(D\). At small bond dimensions, a finite nonzero value of \(M_s\) is observed due to the explicit symmetry breaking inherent to finite-\(D\) MPS representations. As \(D\) is increased, the staggered magnetization gradually decreases and vanishes in the large-\(D\) limit, consistent with the Mermin-Wagner-Hohenberg theorem, which forbids the spontaneous breaking of the continuous \(\mathrm{SU}(2)\) spin-rotation symmetry in one-dimensional systems.

We can now apply the MPS + MF approach explained in the previous sections. The effective interchain exchange $J_{\perp}$ is obtained from the interchain Hubbard parameters listed in Tab.~\ref{tab:hubbard_parameters} according to Eq.~\ref{eq:Jperp} and is antiferromagnetic for all three materials. The resulting $J_{\perp}$ values are also summarized in 
Tab.~\ref{tab:hubbard_parameters}. We find \(J_{\perp} = 1.67\,\mathrm{meV}\) for \(\mathrm{Sr_2CuO_3}\), with corresponding values of \(0.895\,\mathrm{meV}\) and \(1.16\,\mathrm{meV}\) for \(\mathrm{SrBaCuO_3}\) and \(\mathrm{Ba_2CuO_3}\), respectively. 

\begin{figure}
\centering
\includegraphics[width=0.95\linewidth]{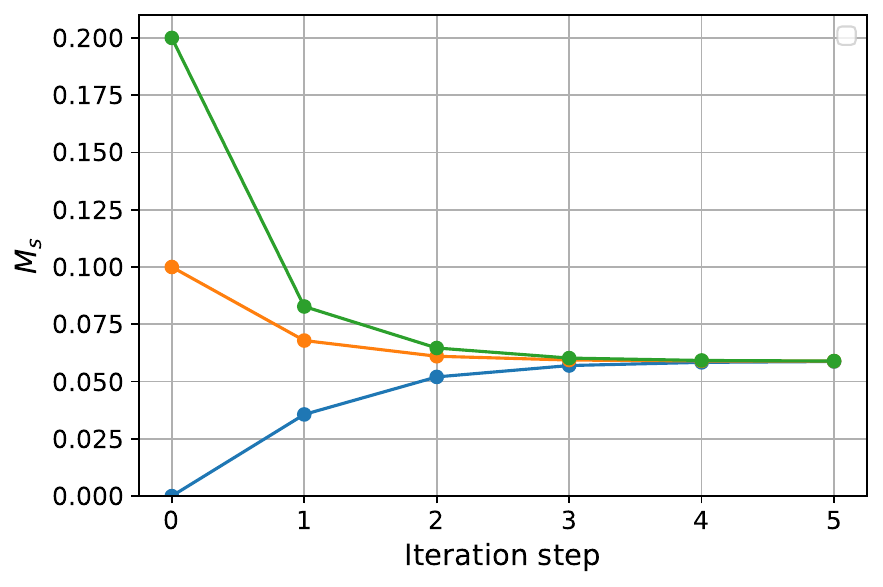}
\caption{Self-consistent convergence of the staggered magnetization \(M_s\) within the MPS + MF framework for \(\mathrm{Sr_2CuO_{3}}\), starting from three different initial guesses \(M_s = 0.2\), \(0.1\) and \(0.0\).}
\label{fig:Ms_vs_iteration}
\end{figure}

Applying the second step of the MPS + MF framework, we introduce an external staggered MF with coupling strength \(J_{\perp}\) and determine the staggered magnetization self-consistently. Starting from an initial guess for \(M_s\), the effective Hamiltonian is solved using MPS, yielding an updated value of \(M_s\), which is then fed back into the MF term. Fig.~\ref{fig:Ms_vs_iteration} illustrates that different initial guesses for the staggered magnetization rapidly converge to the same fixed-point value, demonstrating the insensitivity of the solution to the initial guess.

As shown in Fig.~\ref{fig:Ms_vs_D-Sr2CuO3}, the inclusion of interchain interactions stabilizes a finite, nonzero staggered magnetization, in contrast to the purely one-dimensional case. We find final staggered magnetizations of \(M_s = 0.056\), \(0.045\), and \(0.043\) for \(\mathrm{Sr_2CuO_3}\), \(\mathrm{SrBaCuO_3}\), and \(\mathrm{Ba_2CuO_3}\), respectively. Notably, the value obtained for \(\mathrm{Sr_2CuO_3}\) lies very close to the experimental upper bound of \(M_s \simeq 0.05\) \cite{Ami1995}. Upon substituting Sr with Ba, the staggered magnetization is reduced. However, this reduction cannot be attributed to the interchain exchange $J_{\perp}$, as $J_{\perp}$ itself does not decrease monotonically. This indicates that changes in the intrachain Hubbard parameters also play a crucial role in determining the magnetic order.

Taken together, these results demonstrate that the MPS + MF approach provides a controlled description of the quasi-1D materials by explicitly incorporating interchain couplings. The resulting staggered magnetization of \(\mathrm{Sr_2CuO_3}\) is in quantitative agreement with experimental observation, establishing a direct link between \textit{ab initio}–derived Hubbard parameters and measurable magnetic order in real materials. To our knowledge, experimental estimates of \(M_s\) for \(\mathrm{SrBaCuO_3}\) and \(\mathrm{Ba_2CuO_3}\) are currently lacking, so a direct quantitative comparison is not possible.

\subsection{\label{sec:level2}C. MPS + MF framework for ladders}

In this section, we investigated the extent to which $\mathrm{Sr}_2\mathrm{CuO}_{3.5}$ and $\mathrm{SrCuO}_2$ can be described as ladders.

Based on the Hubbard parameters obtained from the DFT + cRPA downfolding, we solve an effective two-band Hubbard model to compute the ground state, spin gap, and pairing energy of the quasi-1D materials. The hopping and interaction parameters listed in Tab.~\ref{tab:hubbard_parameters2} fully define the Hamiltonian used in the MPS/iDMRG simulations.

\begin{figure}[h]
\centering
\includegraphics[width=0.99\linewidth]{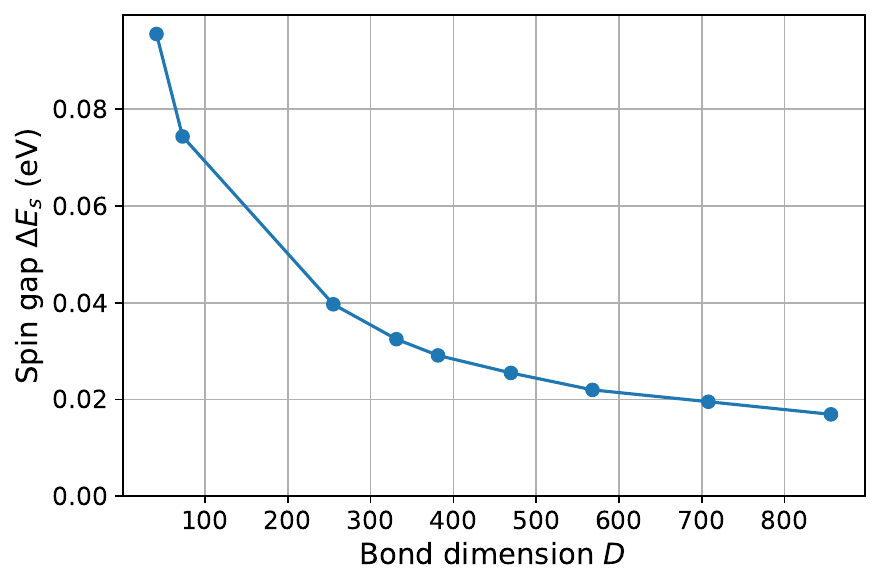}
\caption{The spin gap \(\Delta E_s\) as a function of the bond dimension \(D\) for \(\mathrm{Sr_2CuO_{3.5}}\).}
\label{fig:DeltaS_vs_D}
\end{figure}

Looking at the rung hopping \(t_{\mathrm{rung}}\) for \(\mathrm{SrCuO_2}\), we find it to be extremely small, such that the system is better described as very weakly coupled chains rather than as a two-leg ladder. In the limit of decoupled chains, the one-dimensional repulsive Hubbard model at half filling has a gapless spin sector, $\Delta E_s = 0$, as demonstrated in Refs.~\cite{White1996, Daul1998, Nishimoto2008}.
Hence, for sufficiently small interchain coupling, the spin gap tends to zero as the system approaches the purely one-dimensional limit.
While the rung hopping \(t_{\mathrm{rung}}\) in \(\mathrm{Sr_2CuO_{3.5}}\) exceeds that of \(\mathrm{SrCuO_2}\), the ratio \(t_{\mathrm{rung}}/t_{\mathrm{leg}}\) remains small, \(\sim 0.17\), indicating that the system is still close to the weakly coupled chain regime. Nevertheless, we compute the ground state and evaluate the spin gap \(\Delta E_s\). As shown in Fig.~\ref{fig:DeltaS_vs_D}, the spin gap continues to decrease even at very large bond dimensions, indicating the absence of a robust gap. 

The resulting values of \(\Delta E_s\) are already too small for the perturbative treatment of Bollmark \textit{et al.} \cite{Adrian2023} to be applicable, which requires \(t_\perp \ll 2\Delta E_s\). We therefore conclude that both $\mathrm{Sr}_2\mathrm{CuO}_{3.5}$ and $\mathrm{SrCuO}_2$ lie outside the two-leg ladder regime. As a consequence, the MPS + MF approach of Bollmark \textit{et al.} is not applicable to these systems. However, other materials with a genuine ladder crystal structure may still be well described by a two-leg ladder model.

\section{\label{sec:level1}V. Conclusion and outlook}

In this work, we applied a multi-step workflow for quasi-1D cuprates that combines \emph{ab initio} downfolding (DFT + cRPA) with TNs, while incorporating weak interchain coupling through a self-consistent MF treatment. Starting from relaxed crystal structures, we constructed effective one-band Hubbard models for \(\mathrm{Sr_2CuO_3}\), \(\mathrm{SrBaCuO_3}\), and \(\mathrm{Ba_2CuO_3}\), and solved these models directly in the thermodynamic limit using iDMRG and VUMPS. For an isolated chain, the staggered magnetization decreases systematically as the bond dimension is increased and vanishes in the large-\(D\) limit, which is consistent with the absence of spontaneous \(\mathrm{SU}(2)\) symmetry breaking in 1D at $T=0$ K \cite{Mermin1966, Pitaevskii1991, Tanaka2004}.

Interchain interactions were incorporated through a two-step MF procedure. In the first step, the interchain Hubbard parameters were mapped onto an effective antiferromagnetic Heisenberg exchange \(J_\perp\) using a Schrieffer-Wolff expansion in \(t_\perp/(U - V_\perp)\) \cite{Anderson1950, Anderson1959, Macdonald1988}. In the second step, this exchange interaction was treated within a MF approximation, where each neighboring chain is characterized by its staggered magnetization \(M_s\). This approximation leads to an effective staggered magnetic field acting on a single chain, which can be solved accurately using MPS techniques. The resulting staggered magnetization is determined self-consistently by iterating between the MPS solution and the MF update. The converged ground state exhibits a finite \(M_s\), offering a clear microscopic picture of how weak interchain coupling drives magnetic order beyond the strictly one-dimensional limit.

For the materials \(\mathrm{Sr_2CuO_{3.5}}\) and \(\mathrm{SrCuO_2}\), we assessed the applicability of the perturbative MPS + MF framework of Bollmark \textit{et al.} \cite{Adrian2023} by explicitly computing the intrinsic low-energy scales of isolated ladders. Our results indicate that the intraladder hopping parameters are too small to generate a robust spin gap at half filling: \(\mathrm{SrCuO_2}\) is effectively in the weakly coupled chain regime, while \(\mathrm{Sr_2CuO_{3.5}}\) remains close enough to that limit that \(\Delta E_s\) continues to decrease with bond dimension. Consequently, the separation of scales required for the perturbative interladder treatment is not satisfied, and these materials are better viewed as nearly one-dimensional systems rather than gapped ladders within the present parameterization.

Several natural extensions of this work are possible. On the many body side, incorporating dynamical screening effects and eliminating double counting in the kinetic term, for example with constrained GW based schemes \cite{Hirayama2013}, would improve the quantitative accuracy of the downfolded Hamiltonians. For systems with stronger interchain coupling, a direct treatment using higher dimensional TN approaches such as PEPS \cite{Cirac2021} is a natural next step. Taken together, these directions point toward a practical route for predictive simulations of strongly correlated low dimensional materials that combine controlled \emph{ab initio} model construction with accurate tensor network methods.

\section{\label{sec:level1}Acknowledgments}

We thank Simon Ganne for fruitful discussions. This work was supported by the Research Board of Ghent University (BOF) through a Concerted Research Action (BOF23/GOA/021). V.V.S. acknowledges support from the Research Board of Ghent University (BOF). J.H., D. Verraes, and D. Vrancken acknowledge support from the Fund for Scientific Research–Flanders (FWO) via Grant No. 3G011920, No. 11A3D26N, and No. 11A0B26N, respectively. 
The computational resources (Stevin Supercomputer Infrastructure) and services used in this work were provided by the Flemish Supercomputer Center (VSC), funded by Ghent University, FWO, and the Flemish Government–Department EWI.

\section{\label{sec:level1} Data availability}

The data supporting the findings of this study are included within the article. The input files are available from the authors upon reasonable request. The TN simulations were performed using our in-house Julia code \texttt{HubbardTN.jl}, which is publicly available \cite{HubbardTNrepo}.

\bibliography{apssamp}

\clearpage
\onecolumngrid

\section*{Supporting Information}
\section{\label{sec:level1}I. Orbital projections of upper Hubbard band}

\begin{figure}[h]
\centering
\includegraphics[width=0.5\linewidth]{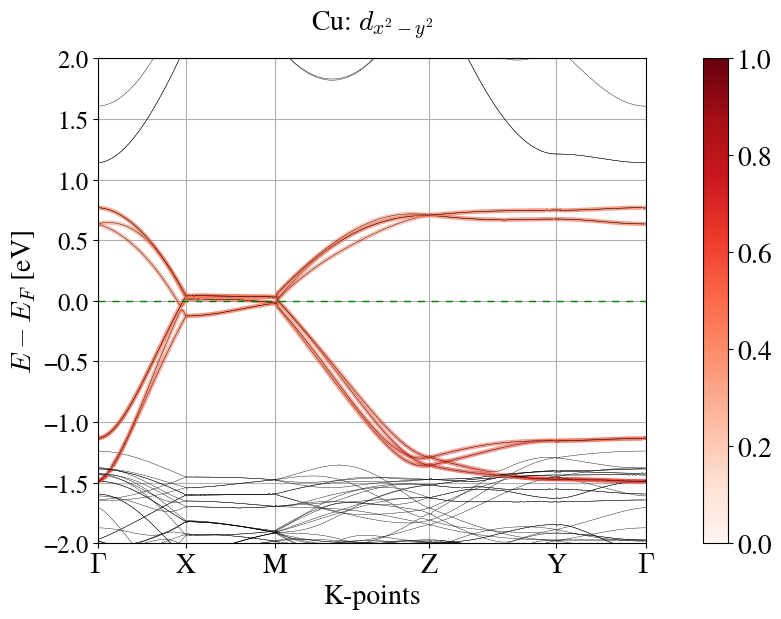}
\caption{The electronic band structure of \( \text{Sr}_2\text{Cu}\text{O}_3 \), highlighting the contribution from the Cu: $3d_{x^2 - y^2}$ orbital. The color scale indicates the orbital character weight, with darker shades corresponding to higher contributions. The high-symmetry points, given in fractional coordinates, are: \(\Gamma = (0, 0, 0)\), \(X = (0.5, 0.0, 0.0)\), \(M = (0.5, 0.0, 0.5)\), \(Y = (0.0, 0.5, 0.0)\), and \(Z = (0.0, 0.0, 0.5)\).}
\label{lm:Sr2CuO3}
\end{figure}

Examining the band structure of \(\text{Sr}_2\text{CuO}_3\), we observe that only a few bands cross the Fermi energy. The orbital contributions of these low-energy states can be analyzed by projecting the calculated KS wavefunctions \(\psi_{nk}^{\text{KS}}\) onto atomic orbitals that are centered on each atom type \(\alpha\) and described by the corresponding spherical harmonics \(Y_{lm}^{\alpha}\). The resulting projection \(|\langle Y_{lm}^{\alpha} | \psi_{nk}^{\text{KS}} \rangle|^2\) yields a decomposition of the initial KS states into contributions from specific angular momentum labeled by \(l\) and \(m\). Most projections onto specific orbitals of O, Cu, or Sr resulted in negligible contributions.
Tab.~\ref{lm:Sr2CuO3-table} lists the dominant orbital contributions to the four bands forming the upper Hubbard manifold. These results clearly indicate that the bands are primarily composed of Cu \(3d_{x^2 - y^2}\) orbitals. This observation is further supported by the \(lm\) decomposition shown in Fig.~\ref{lm:Sr2CuO3}, which highlights the specific bands where the \(3d_{x^2 - y^2}\) character is most significant.

\begin{table}[h]
\centering
\caption{Dominant orbital contributions from an \textit{lm}-decomposition of the upper Hubbard band of \(\text{Sr}_2\text{CuO}_3\), showing the dominant Cu \( d_{x^2 - y^2} \) character. The first column includes the summed $s$ orbital contributions from all atoms (Sr, Cu, and O). Only the mean values over all $k$-points above a threshold of $0.005$ are listed. The total column gives the sum of the squared norms of the wavefunction projections onto atom-centered spherical harmonics within the muffin-tin region. This total is less than 1 due to two effects: (i) the exclusion of wavefunction contributions in the interstitial region (outside the muffin-tin spheres), and (ii) the incompleteness of the projection basis, since spherical harmonics form a complete basis only in the limit \( \ell \to \infty \).}
\label{lm:Sr2CuO3-table}
\begin{tabular}{cccccc}
\hline
\hline
All atoms: $s$ & \(\text{O}2\): $p_y$ & \(\text{O}1\): $p_x$ & Cu: $d_{z^2}$ & Cu: $d_{x^2-y^2}$ & Total \\
\hline
0.021 & 0.202 & 0.171 & 0.007 & 0.481 & 0.882 \\
\hline
\hline
\end{tabular}
\end{table}

\end{document}